\documentclass[fleqn,twoside]{article}
\usepackage{espcrc2}

\usepackage{graphicx}
\usepackage[figuresright]{rotating}

\newcommand{\AmS}{{\protect\the\textfont2
  A\kern-.1667em\lower.5ex\hbox{M}\kern-.125emS}}

\title{From Quarks to Nuclei: 
       Challenges of Lattice QCD}

\author{Sinya Aoki\address{Graduate School of Pure and Applied Sciences, 
        University of Tsukuba, 
        Tsukuba, Ibaraki 305-8571, Japan}%
         \address{Riken-BNL Reserach Center, Brookhaven National Laboratory, Upton, NY 11973, USA}
        }
       
\begin{document}

\begin{abstract}
I discuss challenge of lattice QCD, from quarks to nuclei, which connects QCD with nuclear physics.
\end{abstract}

\maketitle

\section{Introduction}

\begin{figure}[bth]
\centering
\includegraphics[width=65mm,clip]{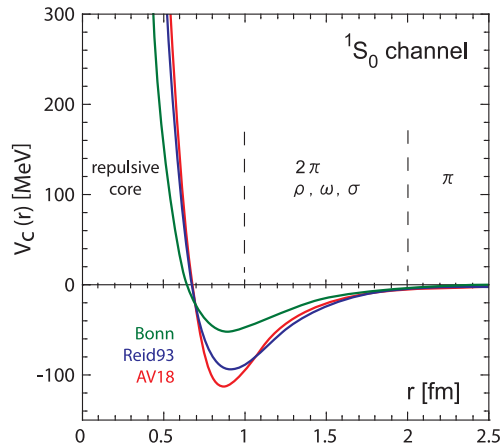}
\hspace*{5mm}
\caption{Three examples of the modern $NN$ potential in $^1S_0$ (spin singlet and $S$-wave) channel: Bonn\cite{Bonn}, Reid93\cite{Reid93} and AV18\cite{AV18}.}
\label{fig:potential}
\end{figure}

 In 1935 Yukawa introduced a virtual particle, now called the pion, to account for the nuclear force\cite{Yukawa},  which bounds protons and neutrons in nuclei. Since then  the nucleon-nucleon ($NN$) interaction has extensively been investigated at low energies both theoretically and experimentally.  Fig.~\ref{fig:potential},  shows modern $NN$ potentials, which are characterized by the following features\cite{Taketani,Machleidt}.  At long distances ($r\ge 2$ fm ) there exists weak attraction, which
is well understood and is dominated by the one pion exchange (OPE).
At medium distances (1 fm $\le r \le $ 2 fm), contributions from the exchange of multi-pions and/or heavy mesons such as $\rho$, $\omega$ and $\sigma$ lead to slightly stronger attraction.
At short distances ($r \le$ 1 fm), on the other hand, attraction turns into repulsion, and it becomes stronger and stronger as $r$ gets smaller, forming the strong repulsive core\cite{Jastrow}.
The repulsive core is essential not only for describing the $NN$ scattering data, but also for the stability and saturation of atomic nuclei, for determining the maximum mass of neutron stars, and for igniting Type II supernova explosions\cite{Supernova}.
Although the origin of the repulsive core must be related to the quark-gluon structure of nucleons,
it remains one of the most fundamental problems in nuclear physics for a long time\cite{OSY}.
It is a great challenge for us to derive the nuclear potential including the repulsive core from lattice QCD.

\section{Potentials from Lattice QCD}
\subsection{Strategy}
\begin{figure}[bth]
\centering
\includegraphics[width=55mm,angle=270,clip]{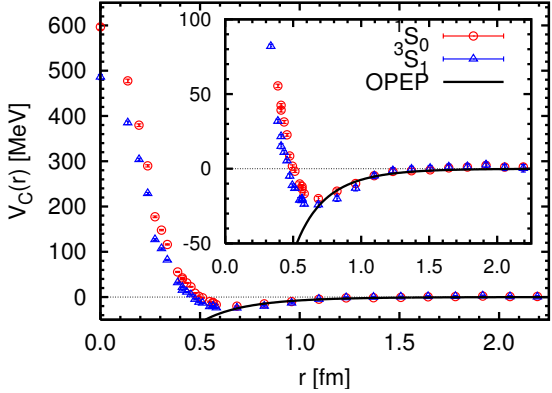}
\caption{The central potential for the singlet  $NN$ (red circles) and the effective central potential for the triplet $NN$ (blue triangles) as a function of $r$ (fm) at $m_\pi\simeq 530$ MeV in quenched QCD\cite{IAH1}. The solid line represents the one pion exchange potential (OPEP) given in Eq.\ref{eq:opep}}
\label{fig:potentialQ}
\end{figure}

Let me explain the strategy proposed in Ref.\cite{IAH1} in order to extract the $NN$ potential from lattice QCD. 

One first defines a wave function with energy $E$ through the equal-time Bethe-Salpeter amplitude as
\begin{eqnarray}
\varphi_E(\vec{r}) &=& \langle 0 \vert N(\vec x,0) N(\vec y, 0)\vert 2N, E\rangle 
\end{eqnarray}
where $\vec r = \vec x -\vec y$, $\vert 2N, E\rangle $ is the 2 nucleon eigen-state of QCD with energy $E$, and $\vert 0 \rangle$ is the vacuum state. More precisely, the 2 nucleon state means a baryon number 2 (or quark number 6) state in QCD.
We here use  a simple local nucleon field operator $N(x)$ given by 
\begin{eqnarray}
N^f_{\alpha}(x) &=& \varepsilon^{abc} (u^a(x) C \gamma_5 d^b(x) ) q^{c,f}_\alpha(x)
\label{eq:nucleon}
\end{eqnarray}
where $a,b,c$ are color, $\alpha$ is the spinor and $f$ is the flavor indices for quark field $q(x) = (u(x), d(x))$, and $C=\gamma_2\gamma_4$ is the charge conjugation matrix in the spinor space.
For large  $r-\equiv\vert \vec r\vert$, it is shown\cite{cp-pacs1} that this wave function behaves as
\begin{eqnarray}
\varphi_l (r) \simeq \frac{A}{k r} \sin( k r - \frac{\pi}{2} l + \delta_l(k))
\end{eqnarray}
where $\varphi_l$ is the $l$-th partial wave of $\varphi$,  $\delta_l(k)$ is the phase shift of the $l$-th partial wave, and $k^2 = 2\mu E = m_N E$ with the reduced (nucleon) mass $\mu$ ($m_N$).

Using the above wave function, the corresponding non-local potential $U(\vec x,\vec y)$ is defined though the
Schr\"odinger equation as
\begin{eqnarray}
[E-H_0] \varphi_E(\vec x) &=& \int d^3 y U(\vec x, \vec y)\varphi_E(\vec y) 
\end{eqnarray}
where $H_0$ is the free Schr\"odinger operator given by
\begin{eqnarray}
H_0&=&\frac{-\nabla^2}{2\mu} .
\end{eqnarray}
Note that $U(\vec x,\vec y$ does {\it not} depend on energy $E$.

Finally one expands $U(\vec x,\vec y)$ in terms of the derivative $\vec \nabla$ as
\begin{eqnarray}
U(\vec x, \vec y)&=&V(\vec x, \vec\nabla)\delta^{(3)}(\vec x-\vec y) .
\end{eqnarray}
Explicitly the leading contribution can be written\cite{OM} as
\begin{eqnarray}
&&V(\vec x, \vec\nabla)=V_0(r) + V_\sigma(r) (\vec\sigma_1\cdot\vec\sigma_2)\nonumber
\\
&+& V_T(r) S_{12} + V_{LS}(r)\vec L\cdot \vec S + O(\nabla^2)
\end{eqnarray}
where $r=\vert \vec x\vert$, $\vec\sigma_{1,2}$ is the spin-operator of each nucleon, 
$\vec L =\vec r\times \vec p$ is the angular momentum operator, $\vec S = \frac{1}{2}(\vec\sigma_1+\vec\sigma_2)$ is the total spin operator, 
and
\begin{eqnarray}
S_{12}&=&\frac{3}{r^2}(\vec\sigma_1\cdot\vec x) (\vec\sigma_2\cdot\vec x)
-  (\vec\sigma_1\cdot\vec\sigma_2)
\end{eqnarray}
is the tensor operator. Coefficient functions $V_X$ has the following decomposition in the flavor space:
\begin{eqnarray}
V_X &=& V_X^0(r)+ V_X^\tau(r) (\vec\tau_1\cdot \vec\tau_2) 
\end{eqnarray}
for $X=0,\sigma,T,LS$, where $\vec\tau_{1,2}$ is the Pauli matrix in the flavor space of each nucleon.

\subsection{First Result}
Applying the strategy mentioned in the previous subsection to the quenched lattice QCD calculation,
the first result of the $NN$ potential has been obtained\cite{IAH1}, as shown in
Fig.\ref{fig:potentialQ}, which reproduces qualitative features of $NN$ potentials such as the strong repulsive core at short distance ( $r\le 0.5$ fm) surrounded by an attraction well at medium and long distances. The solid line in the figure represent the one pion exchange potential given by
\begin{eqnarray}
V_{C}^{\rm OPEP} &=& \frac{g_{\pi N}^2}{4\pi} \frac{(\vec\tau_1\cdot\vec\tau_2)
(\vec\sigma_1\cdot\vec\sigma_2)}{3}\left(\frac{m_\pi}{2m_N}\right)^2\nonumber\\
&\times&  \frac{e^{-m_\pi r}}{r},
\label{eq:opep}
\end{eqnarray}
where $m_\pi \simeq 530$ MeV and $m_N \simeq 1340$ MeV are taken from the simulation, while 
the physical value of the $\pi N$ coupling constant,  $g_{\pi N}^2/(4\pi) \simeq 14.0$, is used. 

\subsection{Questions and Some Answers}
\begin{figure}[bth]
\centering
\includegraphics[width=80mm,clip]{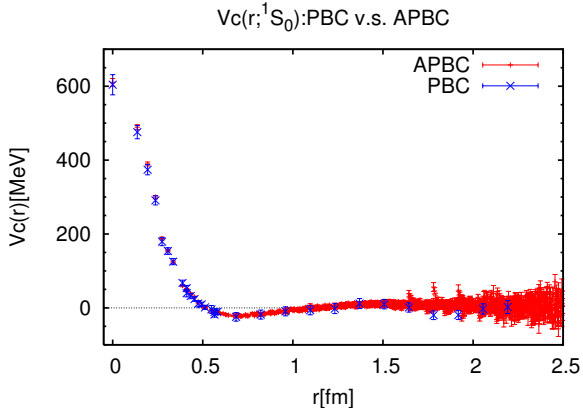}
\caption{The central $NN$ potential for the $^1S_0$ state at $E\simeq 50$ MeV (red bars) and $E\simeq 0$ MeV (blue crosses) in quenched QCD.}
\label{fig:APBC}
\end{figure}So far the strategy to extract $NN$ potentials from lattice QCD seems promising. There exist, however, a few questions to be considered before starting serious investigations of $NN$ or more generally baryon-baryon potential using lattice QCD technique.

As mentioned before, the potential extracted from the wave function is in general energy-independent but non-local:
\begin{eqnarray}
[E-H_0] \varphi_E(\vec x) &=& \int d^3 y U(\vec x, \vec y)\varphi_E(\vec y) .
\end{eqnarray} 
This energy independent non-local potential is equivalent to the local but energy dependent potential,
$V_E$,  defined by
\begin{eqnarray}
[E-H_0] \varphi_E(\vec x) &=& V_E(\vec x)\varphi_E(\vec x) .
\end{eqnarray} 
The 1st question to our strategy is how $V_E(\vec x)$ depends on energy $E$ of the wave function $\varphi_E(\vec x)$. The potential in Fig.\ref{fig:potentialQ}, which is obtained at $E\simeq 0$,
can reproduce the correct phase shift at $E\simeq 0$ (scattering length) by construction. If $V_E(\vec x)$ were strongly dependent on the energy $E$, it would not be so useful: it may not reproduce the correct phase shift at $E^\prime \not= E$. 

The energy dependence of $V_E$ has been investigated in quenched QCD. In Fig.\ref{fig:APBC}
the singlet $NN$ potentials are compared between $E\simeq 0$ MeV and $E\simeq 50$ MeV\cite{murano1}.  For $E\simeq 0$ MeV, the periodic boundary condition (PBC) is used, while the anti-periodic boundary condition (APBC) for $E\simeq 50$ MeV. 
Fluctuations of data with APBC at large distances ( $r\ge 1.5$ fm ) are mainly caused by contaminations from excited states, together with statistical errors. A non-trivial part of the potentials at  $r < 1.5$ fm, on the other hand,  are less affected by such contaminations. As seen from the figure, the $NN$ potentials are almost identical between $E\simeq 0$ MeV and $E\simeq 50$ MeV.
This shows that the potential obtained from our strategy is useful at low energy at least below 50 MeV
at this pion mass ($m_\pi = 530$ MeV) in quenched QCD. It will be important to see if this conclusion holds for lighter pion mass in full QCD.

\vspace{0.5cm}

The second question is the operator dependence of the potential. If the wave function were defined in terms of different $\tilde N(\vec x)$ other than $N(\vec x)$ as
\begin{eqnarray}
\tilde\varphi_E(\vec r)&=& \langle 0 \vert \tilde N(\vec x,0)\tilde N(\vec y,0)\vert 2N, E\rangle,
\end{eqnarray}
one would obtain the different potential:  $\tilde V_E (\vec r) \not= V_E(\vec r)$. 
Our choice of $N(x)$ correspond to a particular choice of the Nucleon effective field theory defined by
\begin{eqnarray}
e^{-S_{\rm eff}(N) }&=&
\int {\cal D}q{\cal D}\bar q {\cal D} A_\mu e^{-S_{\rm QCD}(q,\bar q, A_\mu )}
\nonumber \\
&&\delta\left(
N - \varepsilon^{abc} (u^a C \gamma_5 d^b ) q^{c}\right) .
\end{eqnarray}
Since no field dependence appears at leading order in the Nucleon effective theory,
one expects that operator dependence is also weak at low energy in QCD.
This expectation, however, will have to be explicitly checked in numerical simulations.

\section{Challenges of lattice QCD: Form Quarks to Nuclei }
\begin{figure}[bth]
\centering
\includegraphics[width=55mm,angle=270,clip]{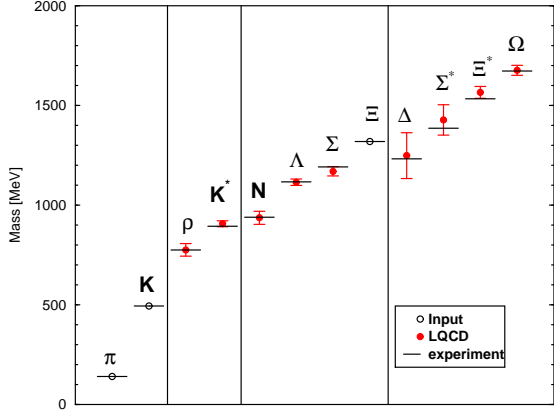}
\caption{Hadron spectra in the continuum limit, together with experimental values.
To fix free parameters of QCD, $m_\pi$, $m_K$ and $m_\Xi$ are used.}
\label{fig:BMW}
\end{figure}
Now I  list some challenges of lattice QCD, which are relevant in nuclear physics.
\begin{enumerate}
\item Full QCD calculations at physical quark masses are definitely important and one of the most difficult challenges in lattice QCD. Once full QCD configurations are generated at physical quark mass
in sufficiently large boxes, one can try to investigate several interesting quantities in nuclear physics. 
For example, one can construct a deuteron on the lattice, or can extract realistic $NN$ potentials from lattice QCD. One can also make real predictions for hyperon interaction from lattice QCD.

\item Calculations of more complicated quantities are also challenges in lattice QCD.
Not only the central part of potentials but also more complicated structures of potentials such as the tensor force or the $\vec L\cdot \vec S$ force should be extracted from lattice QCD. The more advanced challenge in this category is to determine the 3 body force among nucleons, which can hardly be determined experimentally.   Although one can extract the 3 body force from scattering states, it is also a great challenge to construct the bound state of 3 nucleons such as Triton or Helium 3 on the lattice. The lattice QCD may supply inputs for determinations of nuclei structures and inputs for the equation of state inside Super Nova or Neutron Star.

\item Finally we still need some theoretical understandings for the repulsive core of the $NN$ potential.
\end{enumerate}

In the following I will report some activities related to challenges in lattice QCD listed above.

\subsection{Full QCD simulations and hadron spectra}
\begin{figure}[tbh]
\centering
\includegraphics[width=65mm,angle=270,clip]{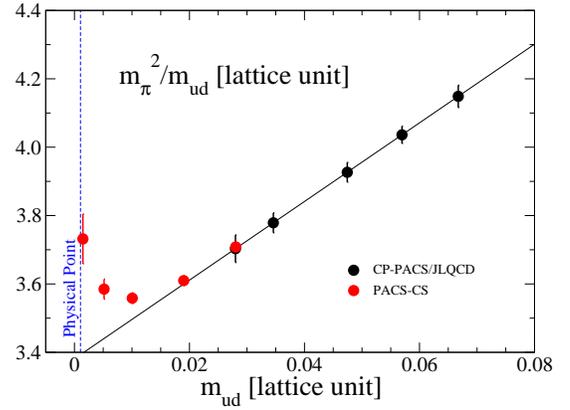}
\caption{The ratio $m_\pi^2/m_{ud}$ as a function of $m_{ud}$ at $a=0.09$ fm and $L=2.9$ fm. Red circles correspond to $m_\pi =$ 702, 570, 412, 296, 156 MeV from right to left. The vertical dashed line is the physical point.}
\label{fig:mpi}
\end{figure}
Fig.\ref{fig:BMW} shows the result of hadron spectra from one of the most extensive 2+1 flavor full QCD calculation
in Ref.\cite{BMW}, where both chiral and continuum extrapolations have been taken. The minimum pion mass in this calculation is $m_\pi^{\rm min} = 190$ MeV, and $m_\pi L \ge 4$ is always satisfied for the spatial extension $L$. The overall scale, the light quark mass and the strange quark mass are fixed by $m_\Xi$, $m_\pi$ and $m_K$. The agreement between lattice QCD and experiment is excellent, as can be seen from the figure.

It is also reported that the pion mass can be smaller than 160 MeV in the 2+1 flavor full QCD
simulations at one lattice spacing $a=0.09$ fm\cite{pacs-cs}. Fig.\ref{fig;mpi} shows the pion mass squired divided by the $ud$ quark mass, $m_\pi^2/m_{ud}$, as a function of $m_{ud}$, where black solid circles are previous results from CP-PACS/JLQCD collaborations\cite{cppacs-jlqcd} 
obtained at heavier quark masses, while red solid circles are new ones from Ref.\cite{pacs-cs}.
As quark mass decreases, this ratio increasingly deviates from the straight line obtained from the fit with black circles. This behavior suggests an existence of the chiral  logarithm predicted by the chiral perturbation theory. Note however that $m_\pi^{\rm min.} L = 2.3$ in this calculation, where the finite size effect might be sizable at the lightest quark mass. To reduce the magnitude of the finite size effect less than a \% level, simulations on larger volume, $L = 5.8 $ fm,  with $m_\pi\simeq 140$ MeV
($ m_\pi L \simeq 4.1$ ), are begun by this group. This simulation indeed will realize the real QCD calculation.
\begin{figure}[tbh]
\centering
\includegraphics[width=55mm,angle=270,clip]{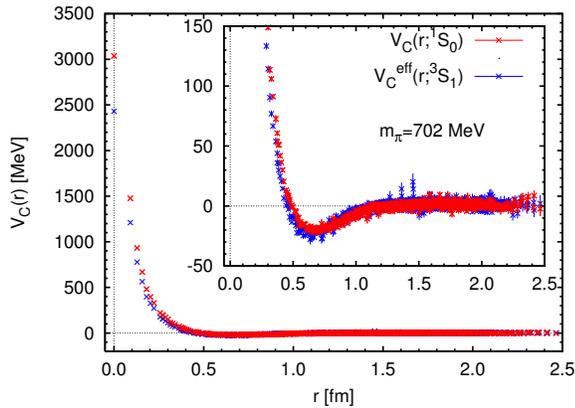}
\caption{2+1 flavor full QCD results of (effective) central potentials at $m_\pi = 702$ MeV.}
\label{fig:NNpotential_full}
\end{figure}

\subsection{Nucleon potential from full QCD simulations}
Using PACS-CS 2+1 flavor full QCD configurations, the $NN$ potentials are currently being calculated.
An example is given in Fig.\ref{fig:NNpotentail_full}, where thesingle central potential $V_C(r; ^1S_0)$
and the triplet effective central potential $V_C^{\rm eff}(r; ^3S_1)$ are plotted as a function of $r$ at $m_\pi = 702$ MeV\cite{IAH2}. By comparing this figure with Fig.\ref{fig:Fig-potential}, one notices that much stronger repulsive core than the quenced case is observed for the potentials in full QCD.
Reason for this difference is currently under investigation.

\subsection{Tensor potential}
\begin{figure}[bth]
\centering
\includegraphics[width=55mm,angle=270,clip]{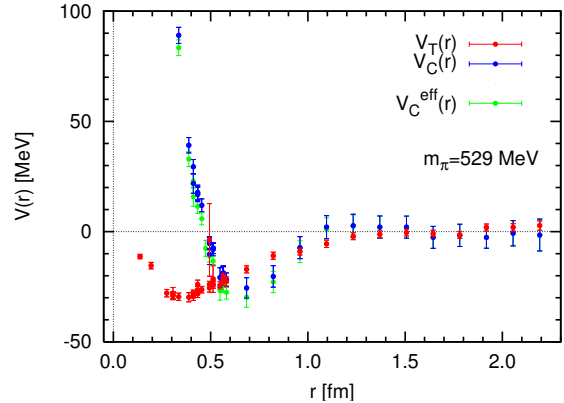}
\caption{The tensor potential $V_T(r)$ (red bars) in quenched QCD at $m_\pi\simeq 530$ MeV, together with the central potentail $V_C(r)$ (blue crosses) and the effective central potential $V_C^{\rm eff} (r)$ (green stars).
}
\label{fig:tensor}
\end{figure}
Since the spin triplet state with $J=1$ contains both $L=0$ and $L=2$, the effective Schr\"odinger equation becomes 
\begin{eqnarray}
\left[E-H_0\right] \varphi_E(\vec r) &=& \left[ V_C(r) + V_T(r) S_{12} \right] \varphi_E(\vec r),
\end{eqnarray}
where
$S_{12}$ is the tensor operator defined before and $V_T(r)$ is the tensor potential.
Using  $P$, the projection operator to $^3S_1$ ( $L=0$ ) state, and
$Q=1-P$, the projection operator to $^3D_1$ ( $L=2$ ) state, the above equation leads to
\begin{eqnarray}
&&\left(\begin{array}{cc}
P\varphi_E(\vec r) & P S_{12}\varphi_E(\vec r) \\
Q\varphi_E(\vec r) & Q S_{12}\varphi_E(\vec r) \\
\end{array}
\right)  \times
\left(\begin{array}{c}
V_C(r) \\
V_T(r) \\
\end{array}
\right) \nonumber \\
&=& (E-H_0)
\left(\begin{array}{c}
P\varphi_E(\vec r) \\
Q\varphi_E(\vec r) \\
\end{array}\right), 
\end{eqnarray}
so that the solution can easily be obtained as
\begin{eqnarray}
\left(\begin{array}{c}
V_C(r) \\
V_T(r) \\
\end{array}
\right)
&=&\left(\begin{array}{cc}
P\varphi_E(\vec r) & P S_{12}\varphi_E(\vec r) \\
Q\varphi_E(\vec r) & Q S_{12}\varphi_E(\vec r) \\
\end{array}
\right)^{-1}  \nonumber \\
& \times &
 (E-H_0)
\left(\begin{array}{c}
P\varphi_E(\vec r) \\
Q\varphi_E(\vec r) \\
\end{array}\right) .
\end{eqnarray}
Note that the $\vec L\cdot \vec S$ potential is neglected here since it is higher order in the derivative expansion.

\begin{figure}[hbt]
\centering
\includegraphics[width=55mm,angle=270,clip]{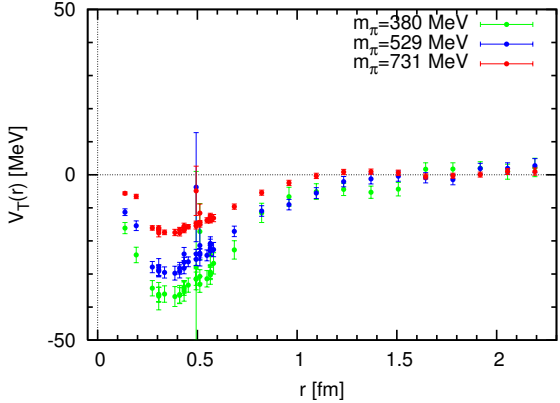}
\caption{The tensor potential at $m_\pi = 380$ MeV (green), 529 MeV (blue) and 731 MeV (red).}
\label{fig:tensor_mass}
\end{figure}

In Fig.\ref{fig:tensor}, the central $V_C(r)$ and the tensor $V_T(r)$ potentials for the spin triplet state, obtained in quenched QCD at $m_\pi \simeq 530$ MeV, are plotted as a function of $r$, together with the effective central potential $V_C^{\rm eff}(r) $. One first notice that $V_{C}(r) \simeq V_C^{\rm eff} (r)$. This means that the neglect of the tensor potential in the previous calculation\cite{IAH1} does not lead to large systematic errors in the determination of the potential.
It is interesting to observe that the tensor potential seems have no repulsive core, and this property is consistence with the estimation to the tensor potential in Ref.\cite{Machleidt2}.
In Fig.\ref{fig: tensor_mass}, $m_\pi$ dependence of the tensor potential is shown.
As seen form the figure, the attraction of the tensor force becomes stronger as the pion mass decreases.
One may expect that the tensor force becomes strong enough at $m_\pi = 135$ MeV to have a bound state, deuteron.

\subsection{Hyperon-Nucleon interactions}
\begin{figure}[htb]
\centering
\includegraphics[width=80mm,clip]{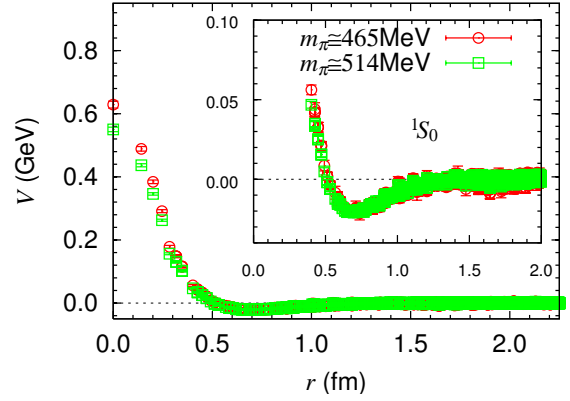}
\caption{The effective central  potential for $\Lambda p$ in the $^1S_0$ channel at $m_\pi \simeq 465$ MeV (red circles) and $m_{\pi}\simeq 514$ MeV (green squares).
 The inset shows its enlargement. }
\label{fig:YN_quench0}
\end{figure}
An important application of the method to extract potential from lattice QCD is to investigate interactions between hyperon-nucleon(YN)  or  hyperon-hyperon(YY), where hyperon(Y) is a baryon which has one or more strange quarks. These interactions, which  are the bases to explore the strange nuclear system,
must be determined experimentally and/or theoretically. Recent systematic study for light $\Lambda$ hypernuclei( $^3_\Lambda$H, $^4_\Lambda$H,  $^4_\Lambda$He and $^5_\Lambda$ He), for example, suggests that the $\Lambda N$ interaction in the $^1S_0$ channel is more attractive than the in the $^3S_1$ channel. The present $YN$ and $YY$ interactions, however, have large uncertainties, since the scattering experiments are either difficult or impossible due to the short life-time of hyperons.
Therefore it is important to determine the $YN$ and $YY$ interactions theoretically from lattice QCD.
\begin{figure}[tbh]
\centering
\includegraphics[width=80mm,clip]{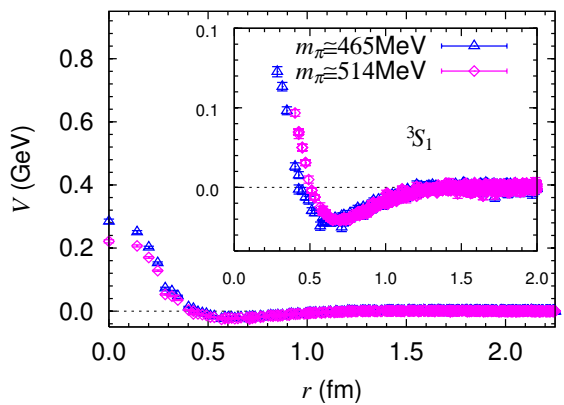}
\caption{
 The effective central potential for $\Lambda p$ in the $^3S_1$ channel
 at $m_{\pi}\simeq 465$ MeV (blue triangles) 
 and $m_{\pi}\simeq 514$ MeV (magenta diamonds).
 The inset shows its enlargement. }
\label{fig:YN_quench1}
\end{figure}

The $N \Lambda$ effective central potentials in quenched QCD are shown in Fig.\ref{fig:YN_quench0}
for the $^1S_0$ state and in Fig.\ref{fig:YN_quench1} for the $^3S_1$ state at $m_\pi = 465$ MeV and 514 MeV\cite{NIAH}. Note that a clear spin dependence of the potential is observed while the pion mass dependence is less significant.

\section{Conclusion}
The challenges of lattice QCD, from quarks to nuclei, have just begun and will continue for a while.
New ideas are important, and close collaboration between particle physics and nuclear physics is
a key for successes of the challenges.
For example, lattice QCD approach to the potential should be complemented by more analytic methods
such as nucleon effective theories\cite{weinberg}.

\section*{Acknowledgements}
First of all, I would like to thank the organizers of this nice winter school for inviting me to give a talk in the panel discussion. This work is supported in part by 
Grants-in-Aid of the Ministry of Education, Culture, Sports, Science and Technology of Japan (Nos. 20340047, 20105001, 20105003).

\end{document}